\documentclass{epl}
\usepackage{epsfig}

\newcommand{\equ}[1]{Eq.~(\protect\ref{#1})}

\newcommand{\bfm}[1]{\mathbf{#1}}

\title{Velocity fluctuations and hydrodynamic diffusion in
sedimentation} 
\shorttitle{Fluctuations in sedimentation} 

\author{M.-Carmen Miguel and R. Pastor-Satorras}

\institute{The Abdus Salam International Centre for Theoretical
Physics (ICTP), P.O. Box 586, 34100 Trieste, Italy}
\pacs{45.70.Qj}{Pattern formation}
\pacs{05.40.-a}{Fluctuation phenomena, random processes, noise, 
          and Brownian motion}
\pacs{82.70.Kj}{Emulsions and suspensions}

\begin{document} 

\maketitle

\begin{abstract} 
  We study non-equilibrium velocity fluctuations in a model for the
  sedimentation of non-Brownian particles experiencing long-range
  hydrodynamic interactions. The complex behavior of these
  fluctuations, the outcome of the collective dynamics of the
  particles, exhibits many of the features observed in sedimentation
  experiments. In addition, our model predicts a final relaxation to
  an anisotropic (hydrodynamic) diffusive state that could be observed
  in experiments performed over longer time ranges.
\end{abstract}

Despite the study of sedimenting suspensions has a long and
well-deserved history for their ubiquitous nature and applications
\cite{Russel95}, attention to the non-equilibrium density and velocity
fluctuations in these systems has only been paid lately.  In
particular, the nature of non-equilibrium fluctuations in the
sedimentation process has been a subject of a long controversy.  While
theoretical arguments~\cite{Caflisch85} and extensive computer
simulations~\cite{Ladd96} suggested that velocity fluctuations should
diverge with the system size, the available experimental
results~\cite{Nicolai95,Xue92}, and the theoretical analysis in
Ref.~\cite{Koch91}, found no evidence for such divergences. These
apparently contradictory observations may have found a reasonable
interpretation after the experimental evidence in Ref.~\cite{Segre97},
and the theoretical study by Levine {\it et al.}~\cite{Levine98}.

Another striking piece in the puzzle of sedimentation was recently
added by the experimental work of Rouyer {\em et al.}~\cite{rouyer99}.
In their experiment, the authors analyzed the trajectories and
velocities of non-Brownian~\cite{note2} particles sedimenting in a
quasi-two dimensional (2d) fluidized bed, and showed the intrinsic
non-Gaussian nature of velocity fluctuations. The main conclusions of
this work are the non-Gaussian form of the probability density
functions (PDF's) of the velocity fluctuations; the anisotropic
character of the particle trajectories (diffusive along the horizontal
direction and superdiffusive along the vertical one); and the presence
of very long-range correlations in the velocity fluctuations along the
gravity direction. New evidence along some of these lines is also
provided in a recent paper by Cowan et al.~\cite{Cowan00}.

The results of Refs.~\cite{Segre97,rouyer99,Cowan00} pose new
questions regarding the process of sedimentation, which have not been
addressed by previous theoretical approaches. Our purpose in this
Letter is to tackle these questions from the point of view of the
particle's dynamics to ascertain the chief physical mechanisms
underlying such fluctuation phenomena.  In order to do so, we propose
a model of sedimentation in which particles experience long-range
hydrodynamic interactions. We start from the solution of the linear
Navier-Stokes equation for the suspension in an unbounded
incompressible fluid~\cite{Russel95}. We consider a system of $N$
particles obeying a system of coupled differential equations which we
solve numerically.  In the solution of the equations, we keep track of
both positions and velocities of the particles, and compute several
relevant statistical properties.  In our model, we observe most of the
experimental features reported in
Refs.~\cite{Segre97,rouyer99,Cowan00}, namely, slow and fast
particles, and swirls and channels in the velocity field, which, in
overall, yield non-Gaussian velocity distributions and a slow time
relaxation of the velocity autocorrelations. In addition, our model
predicts a final relaxation to an anisotropic (hydrodynamic) diffusive
state, not observed in Ref.~\cite{rouyer99}.

The velocity of a particle $n$ in a dilute suspension is given by the
expression ${\bfm U}_n=\sum_m {\bfm H}_{nm}\cdot {\bfm F}_m$, where
the sum is carried out over all the particles $m$ in the suspension,
${\bfm H}_{nm}$ is the mobility tensor, and ${\bfm F}_m$, the external
force acting on each particle, is gravity ${\bfm g}$, oriented along
the positive $z$-axis~\cite{Russel95}.  The simplest form of the
tensor ${\bfm H}$ corresponds to dilute suspensions of point-like
particles.  In this case, the solution of the stationary Navier-Stokes
equation in an unbounded medium yields the so-called Oseen tensor
${\bfm H}_0({\bfm r})=({\bfm I}+{\bfm r}\otimes {\bfm r}/r^2)/8\pi\eta
r$ \cite{note3}, where ${\bfm I}$ is the identity matrix, the operator
$\otimes$ stands for the tensorial product, and $\eta$ is the fluid
viscosity.

We study a suspension of monodisperse non-Brownian particles (for
which inertial effects are irrelevant in a viscous fluid) at very low
concentrations, where the point-particle assumption is indeed a good
approximation. Initially, particles are placed on the same vertical
$xz$ plane.  The form of the mobility matrix in the Oseen
approximation ensures that particles in such a configuration will
never leave that plane.  Simulations are performed on a system of $N$
particles in a square cell of size $L$ (corresponding to a
concentration $c=N/L^2$).  Periodic boundary conditions (PBC's) are
imposed in all directions (including the $y$ direction perpendicular
to the initial plane) in order to guarantee the uniformity of the
suspension \cite{note1}.  To avoid the discontinuities arising from
truncating long-range hydrodynamic interactions, imposing PBC's
amounts to considering Oseen interactions with an infinite set of
images of the original system~\cite{Rapaport95}. In this way, the
velocity of each particle is written as ${\bfm U}_n=\sum_m \sum_{{\bfm
    d}} {\bfm H}_0({\bfm r}_{nm}+{\bfm d})\cdot {\bfm g}$, where the
index $m$ runs through all the particles inside a cell of volume $V$,
${\bfm r}_{nm}$ indicates the relative position of a pair of particles
within that cell, and ${\bfm d}$ runs through the positions of the
images of $m$ in an infinite number of cell replicas along the $x$,
$y$, and $z$ axes.

Imposing PBC's along the $y$ axis is mathematically equivalent to
imposing slip boundary conditions to the fluid velocity field on
effective walls parallel to the sedimentation plane, and located at
distances $\pm L_y/2$, where $L_y<L$.  Sedimentation experiments are
usually carried out within a thin fluid slab confined by parallel
glass plates.  A realistic modelization of this process should thus
include wall effects by imposing no-slip boundary conditions on the
walls. By doing so, hydrodynamic interactions become exponentially
screened for length scales larger than the slab thickness $L_y$.  One
then expects that exponentially damped interactions introduce an
external characteristic length into the problem, $L_y$, which will
govern the dynamics of the system, a fact that has not been pointed
out in the experiments.  On the other hand, short range interactions
severely restrict the extent of the correlations, and render the
dynamics essentially diffusive on all length scales, preventing the
system from showing the collective behavior reported in the
experiments. We have checked this last point by performing simulations
of a system with real walls at distances comparable to the average
interparticle separation.  In particular, with our initial conditions
one obtains a sum of modified Bessel functions which decay
exponentially fast for length scales greater than $L_y$. As expected,
after a short ballistic transient, we observe an essentially diffusive
behavior, quite different indeed from the data reported in
Refs.\cite{Segre97,rouyer99,Cowan00}. We thus conclude that long range
interactions must be preserved in order to account for the scale
competition observed in the system.  Our model is based in this simple
consideration.

To compute ${\bfm U}_n$, we resort to the Ewald summation
method~\cite{Rapaport95}, which yields the following expression:
\begin{eqnarray}\label{eq:2}
  {\bfm M}({\bfm r})&\equiv& \sum_{{\bfm d}} {\bfm H}_0({\bfm r}+{\bfm
    d})\nonumber  \\ 
  &=& \frac{1}{8\pi\eta}\sum_{{\bfm d}} \left\{\frac{{\rm
        erfc}\left(\frac{|{\bfm r}+{\bfm d}|}{2\beta}\right)}{|{\bfm
        r}+{\bfm 
       d}|}\ {\bfm I} 
   +  \left[\frac{{\rm erfc}\left(\frac{|{\bfm r}+{\bfm
             d}|}{2\beta}\right)}{|{\bfm r}+{\bfm
         d}|}+\frac{e^{-\left(\frac{|{\bfm r}+{\bfm
               d}|}{2\beta}\right)^2}}{\sqrt{\pi}\beta} \right]
    \frac{({\bfm 
        r}+{\bfm d})\otimes ({\bfm r}+{\bfm d})}{|{\bfm r}+{\bfm
d}|^2}\right\} \nonumber \\ 
& +& \frac{1}{\eta V} \sum_{{\bfm G}} \frac{e^{i{\bfm G}\cdot {\bfm
r}}\ e^{-\beta^2 G^2}}{G^2}\left[{\bfm I}-(1+\beta^2 G^2)\frac{{\bfm
G} \otimes {\bfm G}}{G^2}\right].
\end{eqnarray} 
Here the function ${\rm erfc}(x)$ is the complementary error function,
and $\beta$ is a parameter which controls the convergence of both the
${\bfm d}$ and ${\bfm G}$ sums. The reciprocal space vectors ${\bfm
G}$ are such that ${\bfm G}\cdot {\bfm d}=2\pi k$, where $k$ is an
integer. The terms proportional to ${\bfm I}$ are the same as for the
Coulomb potential~\cite{Rapaport95}; the other terms are intrinsic to
the Oseen tensor. As in the Coulomb case, the term ${\bfm G}=0$ in
Eq.~(\ref{eq:2}) yields a divergent contribution. In the electrostatic
case, this infinite contribution cancels out after imposing an overall
{\em charge neutrality} condition. In the sedimentation problem, the
${\bfm G}=0$ term cancels out after subtracting the mean
sedimentation velocity ${\bfm U}^M=(N/V) \int d^3 {\bfm r} {\bfm
H}_0({\bfm r}) \cdot{\bfm g}$, and thus working with velocity
fluctuations ${\bfm u}_n={\bfm U}_n-{\bfm U}^M$. By doing so, the
particle positions are on average fixed, as in the sedimentation
experiments in a fluidized bed.

\begin{figure}[t]
\centerline{\epsfig{file=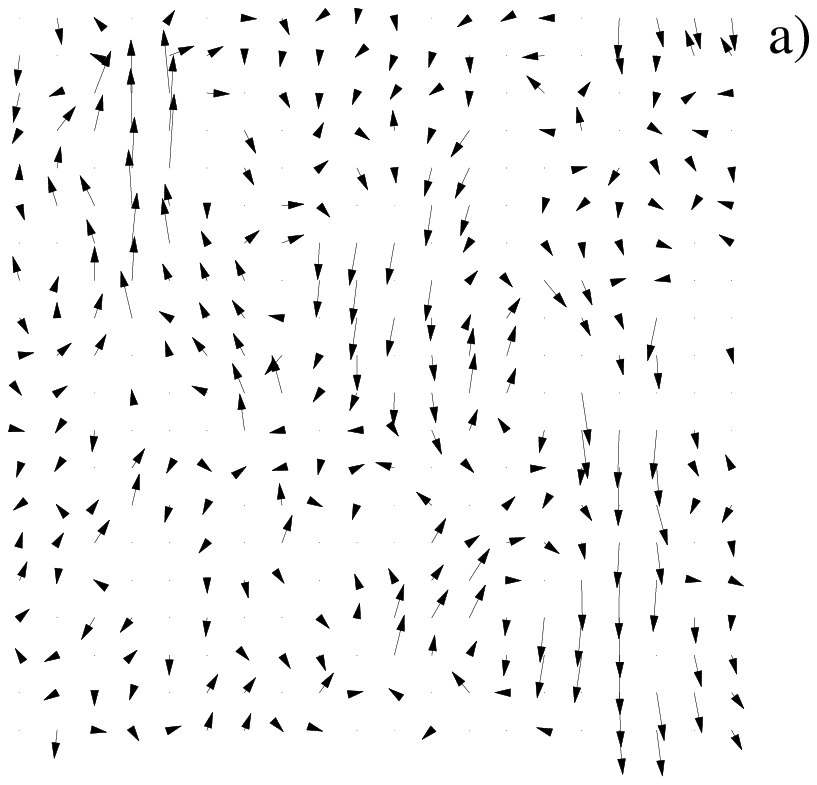, height=6truecm}
\epsfig{file=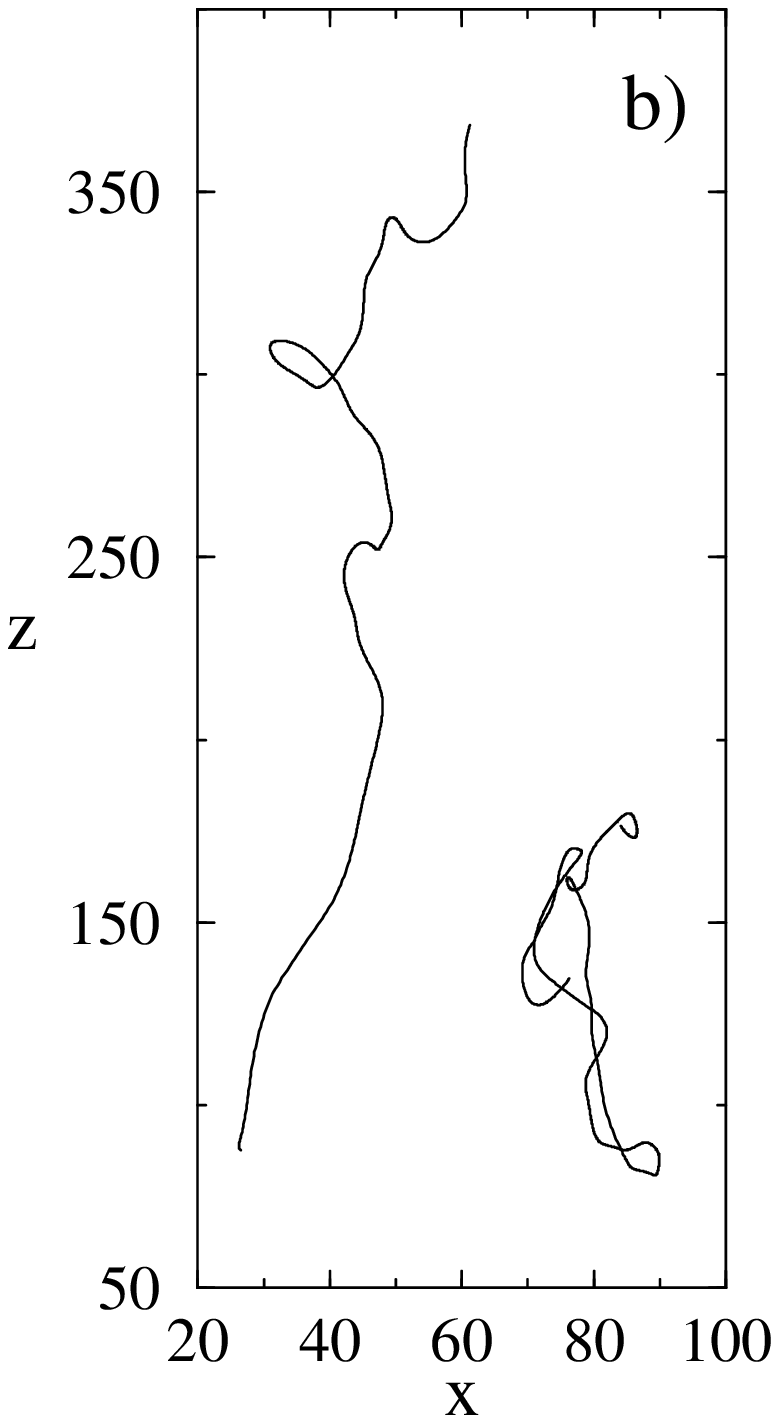, height=6truecm}} 
\caption{a) Snapshot of the velocity fluctuations, showing both swirls
and channels. b) Trajectories of a fast (left) and a slow (right)
particles (see text). Units given in particle radii.}
\label{vortices}
\end{figure}

To follow the evolution of the trajectories and velocities of $N$
particles, we integrate numerically the $2N$ coupled equations $d {\bfm
r}_n / dt= \sum_m {\bfm M}({\bfm r}_{n}-{\bfm r}_{m})\cdot {\bfm
g}-{\bfm U}^M$, where ${\bfm M}$ is given by \equ{eq:2}, using an
adaptive step-size fifth-order Runge-Kutta algorithm~\cite{NumRec}. We
have chosen a convergence parameter $\beta=L/12$; other values of
$\beta$ were also tested, yielding equivalent results.  Simulations
start from a configuration of $N$ particles randomly placed on a
square cell of size $L$.  Since the Oseen approximation is not valid
at short distances, to avoid singularities in the velocity field we
have introduced an {\em ad hoc} very short range repulsive hard-core
term of the form $\exp[-(r-2a)/\rho]$, where $a$ is the radius of the
particles and $\rho$ is a small parameter that we select equal to
$0.1$.

The concentrations described by our model are severely limited by both
the range of validity of the Oseen approximation and the available CPU
time.  In our simulations, therefore, we have considered
concentrations $c\leq 1\%$, and cell sizes ranging from $L=100a-200a$.
We shall see, however, that our results for dilute concentrations
already exhibit most of the salient features reported in the
literature.  Averages were made over at least $100$ realizations
starting with different random initial conditions.

In the evolution of our model, particles build up a complex and highly
fluctuating pattern of velocity swirls and channels, very similar to
those experimentally observed~\cite{Segre97,rouyer99}. In
Fig.~\ref{vortices}a), we show a snapshot of a system with
concentration $c=1\%$ and cell size $L=200a$.  The number of swirls
and their sense of flow (clock- or anti-clock-wise) result from the
collective interactions, and have the constraint of zero global
vorticity $\sum_n \nabla \times {\bfm u}_n=0$, as follows from the
symmetries of the Oseen tensor.

\begin{figure}[t]
  \centerline{\epsfig{file=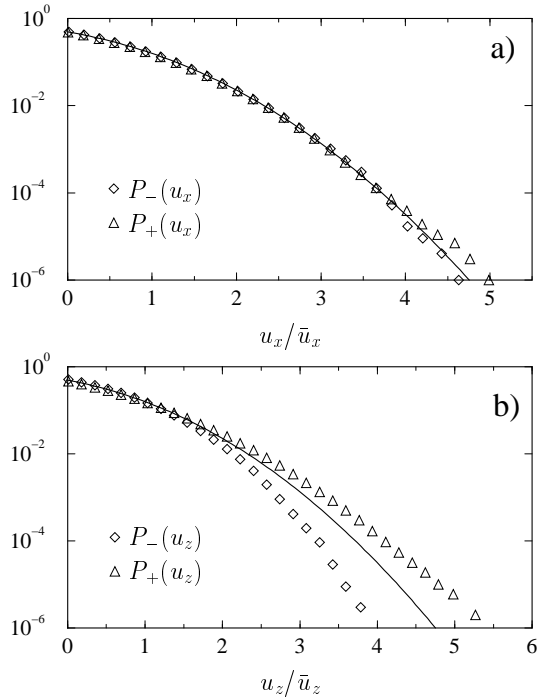, width=7truecm}}
  \caption{Integrated distributions $P_+$ and $P_-$ for the (a) $u_x$
  and (b) $u_z$ velocity fluctuations in linear-log scale. The solid
  line corresponds to an integrated Gaussian distribution.}
  \label{pdvs}
\end{figure}

At large times, the average root-mean-square velocity fluctuations
(RMSVF) along the horizontal, $\bar{u}_x$, and vertical directions,
$\bar{u}_z$, grow with the concentration $c$.  As naively expected
from the symmetry breaking induced by gravity, fluctuations are
anisotropic.  We measure a ratio $\bar{u}_z/\bar{u}_x\simeq2.5$, which
seems independent of $c$ or $L$. This observation agrees with the
results reported in Ref.~\cite{Segre97}.

Next, we have measured the probability density function (PDF) of the
velocity fluctuations, $p(u_z)$ and $p(u_x)$, normalized as to have
zero mean and unity standard deviation. In Fig.~\ref{pdvs} we plot the
integrated distribution functions $P_+(u) = \int_u^\infty p(u') du'$
for the downward (rightward), $u>0$, velocity, and $P_-(-u) =
\int_{-\infty}^u p(u') du'$ for the upward (leftward), $u<0$,
velocity, for both vertical and horizontal fluctuations. In
particular, the plots correspond to values of $c=1\%$ and $L=200a$.
We observe that the horizontal fluctuations are left-right symmetric
and very well approximated by a Gaussian distribution (solid line in
Fig.~\ref{pdvs}a)).  On the other hand, vertical fluctuations are
fairly asymmetric and apparently non-Gaussian.

\begin{figure}[t]
\centerline{\epsfig{file=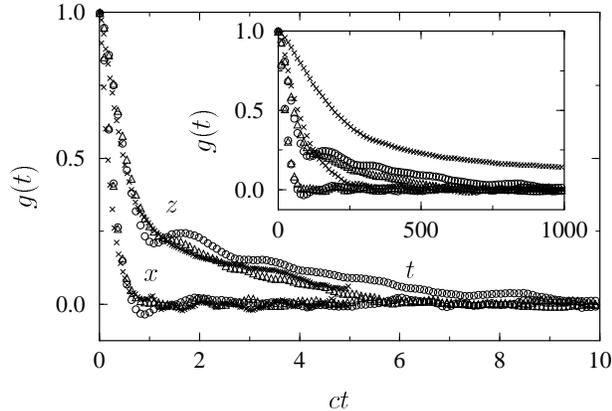, width=8truecm}} 
\caption{Velocity autocorrelations as a function of time.  The curves
shown in the inset correspond to the raw data, whereas in the main
plot, time has been rescaled by the characteristic time $\tau \sim
c^{-1}$. ($\circ$) system I, ($\times$) system II, ($\triangle$)
system III (see text).}  
\label{gdt}
\end{figure}

We now turn our attention to the two-times statistical properties of
the velocity fluctuations. First, we consider the velocity
autocorrelation function $g_{\alpha}(t)=\left<u_{\alpha}(0)
  u_{\alpha}(t)\right>/\left<u_{\alpha}(0)^2\right>$, for
$\alpha=x,z$, where the brackets denote an average over particles and
realizations, at a fixed time $t$. In Fig.~\ref{gdt} we depict
$g_{\alpha}(t)$ for two different concentrations, $c=1\%$ (system I)
represented with ($\circ$), and $c=0.25\%$ (system II) represented
with ($\times$), in a box of size $L=200$, as well as $c=1\%$ in a
smaller box of size $L=100$ (system III) which we plot with
($\triangle$).  The main plot represents our data as a function of the
rescaled time $c t$; raw data are shown in the inset.  For both $g_x$
and $g_z$, we observe an initial exponential decay of the correlations
with a characteristic time proportional to $c^{-1}$.  This scaling of
$g_\alpha$ at short times can be understood by means of a simple
mean-field-like argument: Given the expression of the Oseen tensor,
the velocity correlations can be written as $\left<u(t) u(0)\right>
\sim \left<u(0)/r(t) \right>$, where $r$ is the separation between any
pair of particles.  Taking a time derivative, $\partial_t \left<u(t)
  u(0) \right> \sim \partial_t \left<u(0)/r(t) \right> \sim - \left<
  u(t) u(0)/ r^2\right>$, where in the second step we have commuted
derivative and average. A further simplification considers $r\sim
c^{-1/2}$, i.e., the average separation between particles. Then, we
have $\partial_t \left<u(t) u(0)\right>\sim -
\left<u(t)u(0)\right>/c^{-1}$, yielding an  exponential
relaxation with characteristic time $\tau \sim c^{-1}$.

After this initial decay, the $x$ correlations of the more
concentrated system I show a clear negative region.  Curiously, this
behavior resembles that of a dense liquid. Negative autocorrelations
in a dense liquid are due to backscattering effects after collisions
among molecules. In our system, however, negative correlations are due
to the permanence of the particles in a velocity swirl. As argued
in~\cite{rouyer99}, during the course of a simulation some of the
particles become part of velocity swirls and spend in them a
considerable amount of time.  They can be called {\em slow} particles
and describe coil-like trajectories. Others ({\em fast} particles)
spend more time inside the channels separating swirls, and their
trajectories are much more elongated. Both channels and swirls can be
observed in Fig.~\ref{vortices}a). In Fig.~\ref{vortices}b), we plot
typical trajectories of a fast and a slow particles.

At later times the correlations of the $x$ components oscillate around
zero, whereas the $z$ autocorrelations go through a second regime of
much slower relaxation, and eventually become zero towards the end of
the simulation time. This enhancement of the $z$ autocorrelations is
due to the very existence of channels between swirls, inside of which
particles follow ballistic trajectories with small fluctuations.
Channels are interrupted by swirls, but since these must be created in
pairs of opposite vorticity (due to vorticity conservation), their
creation is costly and only a few are present in a box of small
size. A long time is thus required for the particles initially in a
channel to become part of a few velocity swirls and uncorrelate from
their initial conditions.

\begin{figure}[t]
  \centerline{\epsfig{file=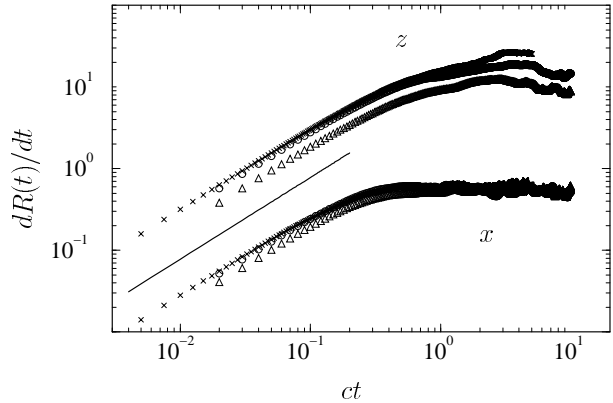, width=8truecm}} 
 \caption{Time
  derivative of the mean-square displacement in a double logarithmic
  scale. The solid line with slope $1$ represents the ballistic
  regime.}  \label{walks}
\end{figure}

To further explore the behavior of the system at long times, we have
also studied the mean-square displacement of the particles,
$R_{\alpha}(t)=\left<[r_{\alpha}(t)-r_{\alpha}(0)]^2\right>$, which is
an efficient indicator of a possible effective diffusive behavior of
the system (hydrodynamic diffusion)~\cite{Ladd96}. For the latter, we
expect $R_\alpha(t)=2D_{\alpha} t$, i.e.,
$dR_\alpha(t)/dt=2D_{\alpha}\equiv$ const, where $D_{\alpha}$ is an
effective diffusion coefficient.  In Fig.~\ref{walks} we represent the
time derivative of $R_{\alpha}(t)$ for the displacements along the $x$
and $z$ directions.  The plateau at long times clearly indicates that
the displacement along the $x$ direction becomes purely diffusive
right after an initial ballistic regime ($R_x(t) \sim t^2)$.  The $z$
displacement also becomes eventually diffusive, but at longer times
scales.  This final diffusive behavior is compatible with the fast
decay  of the tails of the PDF's shown in
Fig.~\ref{pdvs}. We observe that $D_z \gg D_x$, hence the diffusive
regime is highly anisotropic.  At intermediate times, we observe that
$R_z(t)$ can be fitted to a power-law $R_z(t)\sim t^{\alpha}$ with an
exponent within the range $1-2$. Such behavior was reported in
\cite{rouyer99}, where experiments could not be run for long enough
times as in our simulations. We expect that experiments carried out
over longer time scales would also show the eventual diffusive
behavior along the vertical direction.

To sum up, we present a model for the sedimentation of non-Brownian
particles in an unbounded fluid that incorporates long range
hydrodynamic interactions and PBC's in the simplest Oseen
approximation.  This model exhibits most of the salient features of
the experiments reported in Refs.~\cite{Segre97,rouyer99,Cowan00}.
Our findings can be understood within the picture of slow and fast
particles: Slow particles spend most of the time within velocity
swirls and contribute to the fast relaxation of the velocity
correlations. Fast particles moving along velocity channels have
strongly correlated (quasi-ballistic) trajectories and are responsible
for the slow relaxation component.  For sufficiently long times, all
particles become part of enough velocity swirls, and our model
predicts that the system eventually relaxes to a hydrodynamic
diffusive regime, that could be confirmed by experiments performed
over longer time spans. 

\vspace*{0.25cm}
\centerline{***}
\vspace*{0.25cm}

We thank \Name{S. Franz, M. Kardar, I. Pagonabarraga, M. Rub\'{\i}, \and
A. Vespignani} for helpful discussions. The work of R.P.S. has been
supported by the TMR Network ERBFMRXCT980183.


\begin{thebibliography}{99}
\bibitem{note2}
  Particles for which the diffusive motion caused by the
  surrounding solvent is negligible.
\end{thebibliography}
\end{document}